\documentclass[prc,english,showpacs,showkeys,superscriptaddress,
nofootinbib,preprintnumbers,amsmath,amssymb,fleqn,byrevtex]{revtex4}
\usepackage{amsmath,amsfonts,amssymb,amscd,amsxtra,amsthm,amstext}
\usepackage[T1]{fontenc}
\usepackage[latin9]{inputenc}
\usepackage{bm}
\usepackage{graphicx}
\usepackage{esint}
\usepackage{dcolumn}
\usepackage{babel}

\begin{document}

\preprint{INHA-NTG-02/2012}

\title{$\Theta^+$ baryon, $N^* (1685)$ resonance, and $\pi 
  N$ sigma term \\ in the context of the LEPS and DIANA experiments}    
\author{Ghil-Seok Yang}
\email{ghsyang@knu.ac.kr}
\affiliation{Center for High Energy Physics and Department of Physics, 
  Kyungpook National University, Daegu 702-701,  Republic of Korea}

\author{Hyun-Chul Kim}
\email{hchkim@inha.ac.kr}
\affiliation{Department of Physics, Inha University, Incheon 402-751,
  Republic of Korea}
\affiliation{Department of Physics, University of Connecticut,
  Storrs, CT 06269, U.S.A.} 
\affiliation{School of Physics, Korea Institute for Advanced Study,
  Seoul 130-722, Republic of Korea}

\date{April, 2012}

\begin{abstract}
We reexamine properties of the baryon antidecuplet $\Theta^+$ and
$N^*$, and the $\pi N$ sigma term within the framework of a chiral
soliton model, focusing on their dependence on the $\Theta^+$ mass.
It turns out that the measured value of the $N^*$ mass,
$M_{N^*}=1686$ MeV, is consistent with that of the $\Theta^+$ mass 
$M_{\Theta^+}=1524$ MeV by the LEPS collaboration. The $N^*\to
N\gamma$ magnetic transition moments are almost independent of the
$\Theta^+$ mass. The ratio of the radiative decay width
$\Gamma_{nn^*}$ to $\Gamma_{pp^*}$ turns out to be around 5. The decay
width for $\Theta^+\to NK$ is studied in the context of the LEPS and
DIANA experiments. When the LEPS value of the
$\Theta^+$ mass is employed, we obtain $\Gamma_{\Theta NK}
=(0.5\pm0.1)$ MeV. The $\pi N$ sigma term is found to be almost 
independent of the $\Theta^+$ mass. In addition, we derive a new
expression for the $\pi N$ sigma term in terms of the isospin mass
splittings of the hyperon octet as well as that of the antidecuplet $N^*$.       
\end{abstract}


\maketitle

\section{Introduction}
The baryon antidecuplet is the first excitation consisting
of exotic pentaquark
baryons~\cite{Jezabek:1987ns,Diakonov:1997mm,Praszalowicz:2003ik}.      
Since the LEPS collaboration reported the first measurement of the
pentaquark baryon $\Theta^{+}$~\cite{Nakano:2003qx}, the pentaquark 
baryons have attracted much attention, before a series of the CLAS
experiments has announced null results of the $\Theta^{+}$
~\cite{Battaglieri:2005er,McKinnon:2006zv,Niccolai:2006td,
DeVita:2006ha}, which casted doubt on the 
existence of the pentaquarks~\cite{Close:2005pm,wohl}. On the other  
hand, the DIANA collaboration has pursued searching for the 
$\Theta^{+}$~\cite{Barmin:2006we,Barmin:2009cz} and observed the 
formation of a narrow $pK^{0}$ peak with mass of $1538\pm2$
MeV/$c^{2}$ and width of $\Gamma=0.39\pm0.10$ MeV in the
$K^{+}n\rightarrow K^{0}p$ reaction with higher statistical 
significance ($6\,\sigma-8\,\sigma$)~\cite{Barmin:2009cz}. 
The decay width was more precisely measured in comparision with the
former DIANA measurement~\cite{Barmin:2003vv}, the statistics being
doubled.  The SVD experiment has also found a narrow peak with the
mass, $(1523\pm2_\mathrm{stat.}\pm3_\mathrm{syst.})$ MeV in the
inclusive reaction $pA\to pK_s^0+X$~\cite{SVD,SVD2:2008}. 
In 2009, the LEPS collaboration has again announced the evidence 
of the $\Theta^+$~\cite{Nakano:2008ee}: The mass of the $\Theta^+$ is
found at $M_{\Theta}=1524\pm2\pm 3\,\mathrm{MeV}/c^{2}$ and the
statistical significance of the peak turns out to be $5.1\,\sigma$.
The differential cross section was estimated to be $(12\pm2)\,
\mathrm{nb/sr}$ in the photon energy ranging from 2.0 GeV to 2.4 GeV
in the LEPS angular range. Note that the statistics of the new LEPS
data has been improved by a factor of 8 over the previous
measurement~\cite{Nakano:2003qx}. Very recently, Amaryan et al. have
reported a narrow structure around 1.54 GeV in the process
$\gamma+p\to pK_SK_L$ via interference with $\phi$-meson production
with the statistical significance $5.9\,\sigma$, based on the CLAS
data~\cite{Amaryan:2011qc}. 

In addition to the $\Theta^{+}$ baryon, Kuznetsov et
al.~\cite{Kuznetsov:2006kt} have observed a new nucleon-like resonance
around $1.67$ GeV from $\eta$ photoproduction off the deutron in the
neutron channel. The decay width was measured to be around $40$
MeV without the effects of the Fermi motion~\cite{Fix:2007st} excluded.
On the other hand, this narrow resonant structure was not seen
in the quasi-free proton channel~\cite{Kuznetsov:2006kt}. The finding
of Ref.~\cite{Kuznetsov:2006kt} is consistent with the theoretical 
predictions~\cite{Diakonov:2003jj,Arndt:2003ga} of non-strange exotic  
baryons. Moreover, its narrow width and isospin asymmetry in the
initial states, also called as the neutron
anomaly~\cite{Kuznetsov:2008ii}, are the typical characteristics for
the photo-excitation 
of the non-strange antidecuplet
pentaquark~\cite{Polyakov:2003dx,Kim:2005gz}.   
New analyses of the free proton GRAAL
data~\cite{Kuznetsov:2007dy,Kuznetsov:2008gm,  
Kuznetsov:2010as,Kuznetsov:2008hj,Kuznetsov:2008ii}  
have revealed a resonance structure with a mass around 1685 MeV and
width $\Gamma\leq15$ MeV, though the data of
Ref.~\cite{Bartalini:2007fg} do not agree with those of
Ref.~\cite{Kuznetsov:2007dy}. For a detailed discussion of this 
discrepancy, we refer to Ref~\cite{Kuznetsov:2008gm}. 
The CB-ELSA collaboration~\cite{CBELSA,Jaegle:2008ux,Jaegle:2011sw}
has also confirmed an evidence for this $N^{*}$ resonance in line with
those of GRAAL. Very recently, Kuznetsov and Polyakov have
extracted the  new result for the narrow peak: $M_{N^*}=1686\pm 7\pm
5$ MeV with the decay width $\Gamma\approx 28\pm 12$
MeV~\cite{Kuznetsov:2011zz}. All these experimental facts are 
compatible with the results for the transition magnetic moments in the   
chiral quark-soliton model 
($\chi$QSM)~\cite{Polyakov:2003dx,Kim:2005gz} and phenomenological
analysis for the non-strange pentaquark  
baryons~\cite{Azimov:2005jj}. the $\gamma N\to\eta N$ reaction was
studied within an effective Lagrangian 
approach~\cite{Choi:2005ki,Choi:2007gy} that has described
qualitatively well the GRAAL data. The present status of the
$N^*(1685)$ is summarized in Ref.~\cite{Polyakov:2011qr} in which the
reason was discussed why the $N^*(1685)$ can be most probably
identified as a member of the baryon antidecuplet in detail.

In the present work, we want to examine the relation between the
$\Theta^+$ mass and other observables such as the mass of the $N^*$
($M_{N^*}$), $N^*\to N\gamma$ transition magnetic moments
($\mu_{NN^*}$), the decay width of the $\Theta^+$
($\Gamma_{\Theta^+}$), and $\pi N$ sigma term ($\sigma_{\pi N}$),  in 
the context of the LEPS and DIANA experiments. In particular, we will
regard the $N^*(1685)$ resonance with the narrow width as a member of
the antidecuplet in this work. The mass splittings of the SU(3)
baryons within a chiral soliton model ($\chi$SM) were 
reinvestigated with all parameters fixed
unequivocally~\cite{Yang:2010fm}. Since the mass of the $\Theta^+$
observed by the LEPS collaboration is different 
from that by the DIANA collaboration, it is of great importance to examine
carefully the relevance of the analysis in Ref.~\cite{Yang:2010fm}
with regard to the LEPS and DIANA experiments.   
We will show in this work that the decay width  
$\Gamma_\Theta$ obtained from the $\chi$SM~\cite{Yang:2010fm}
is consistent with these two experiments. We will also study the
dependence of the $N^*$ mass on the $M_\Theta$, which turns out to be
compatible with the LEPS data. In addition, we also investigate the
dependence of the $N^*\to N$ magnetic transition moment that 
is shown to be almost insensitive to the $\Theta^+$ mass. Finally,  
the $\sigma_{\pi N}$ will be examined, which becomes one of essential
quantities in the physics of dark 
matter~\cite{Ellis:2008hf,Ellis:2012aa}. Motivated by its relevance in
dark matter, a great amount of efforts was put on the evaluation of
the $\sigma_{\pi N}$. For example, there are now various results from
the lattice
QCD~\cite{Leinweber:2003dg,Ohki:2008ff,Durr:2011mp,Bali:2011ks}.  
However, the value of $\sigma_{\pi N}$ still does not converge, but is  
known only with the wide range of uncertantities: $35-75$ MeV.
Thus, we will discuss the $\sigma_{\pi N}$ in connection with the
baryon antidecuplet and will show that it is rather stable with
respect to the $\Theta^+$ mass. Moreover, its predicted value is
smaller than that used in previous
analyses~\cite{Diakonov:1997mm,Ellis:2004uz,Schweitzer:2003fg}.  

The present work is organized as follows: In Section II, the pertinent
formulae for the baryon antidecuplet within a chiral soliton model are
compiled.  In Section III, we discuss the results. Final Section is
devoted to summary and conclusion.

\section{Baryon antidecuplet from a chiral soliton  model}  
We first recapitulate briefly the formulae of
the mass splittings, the magnetic moments, and the axial-vector
constants within the framework of the $\chi$SM. We begin with the 
collective Hamiltonian of chiral solitons, which have been thoroughly
studied within various versions of the $\chi$SM 
such as the chiral quark-soliton
model~\cite{Blotz:1992pw,Blotz:1992br}, the Skyrme
model~\cite{Weigel:2008zz}, and the chiral hyperbag
model~\cite{Park:1988vy}.  
The most general form of the collective Hamiltonian in the SU(3) 
$\chi\mathrm{SM}$ can be written as follows: 
\begin{equation}
H \; = \; M_{{\rm cl}}\;+\; H_{\mathrm{rot}}\;+\; H_{\mathrm{sb}},
\label{eq:collH}
\end{equation}
where $M_{\mathrm{cl}}$ denotes the classical soliton mass. The
$H_{\mathrm{rot}}$ and $H_{\mathrm{sb}}$ respectively stand for
the $1/N_{c}$ rotational and symmetry-breaking corrections with
the effects of isospin and $\mathrm{SU(3)}$ flavor symmetry 
breakings included~\cite{Blotz:1994pc}:
\begin{eqnarray}
H_{\mathrm{rot}} 
& = & 
\frac{1}{2I_{1}}\sum_{i=1}^{3}\hat{J}_{i}^{2}
\;+\;\frac{1}{2I_{2}}\sum_{p=4}^{7}\hat{J}_{p}^{2},
\label{eq:rotH}\\
H_{\mathrm{sb}}
& = & 
\left(m_{\mathrm{d}}-m_{\mathrm{u}}\right)\left(\frac{\sqrt{3}}{2}\,
  \alpha\, D_{38}^{(8)}(A) \;+\;\beta\,\hat{T_{3}}
\;+\;\frac{1}{2}\,\gamma\sum_{i=1}^{3}D_{3i}^{(8)}(A)\,
\hat{J}_{i}\right)\cr 
 &  & +\;\left(m_{\mathrm{s}}-\bar{m}\right)\left(\alpha\,
   D_{88}^{(8)}(A)
\;+\;\beta\,\hat{Y}
\;+\;\frac{1}{\sqrt{3}}\,\gamma\sum_{i=1}^{3}D_{8i}^{(8)}(A)
\,\hat{J}_{i}\right)\cr  
 &  & -\;\left(m_{u}+m_{d}+m_{s}\right) (\alpha+\beta),
\label{eq:sbH}
\end{eqnarray}
where $I_{1,2}$ represent the soliton moments of inertia that depend
on dynamics of specific formulations of the $\chi$SM. The $J_{i}$
denote the generators of the SU(3) group. The $m_{\mathrm{u}}$,
$m_{\mathrm{d}}$, and $m_{\mathrm{s}}$ designate the up, down, and
strange current quark masses, respectively. The $\bar{m}$ is the
average of the up and down quark masses. The
$D_{ab}^{(\mathcal{R})}(A)$ indicate the SU(3) Wigner $D$
functions. The $\hat{Y}$ and $\hat{T_{3}}$ are the operators of the
hypercharge and isospin third component, respectively. 
The $\alpha$, $\beta$, and $\gamma$ are given in terms of the
$\sigma_{\pi N}$ and soliton moments of inertia $I_{1,2}$ 
and $K_{1,2}$ as follows: 
\begin{equation}
\alpha=-\left(\frac{\sigma_{\pi N}}{3 \bar{m}}-\frac{K_{2}}{I_{2}}\right),
\;\;\;\;\beta=-\frac{K_{2}}{I_{2}},
\;\;\;\;\gamma=2\left(\frac{K_{1}}{I_{1}}-\frac{K_{2}}{I_{2}}\right).
\label{eq:abg}
\end{equation}
 Since $\alpha$, $\beta$, and $\gamma$ depend on the moments of
inertia and $\sigma_{\pi N}$, they are also related to details of
specific dynamics of the $\chi\mathrm{SM}$. Note that $\alpha$,
$\beta$, and $\gamma$ defined in the present work do not contain the
strange quark mass, while those in
Refs.~\cite{Diakonov:1997mm,Ellis:2004uz} include it. 

In the $\chi$SM, we have the following constraint for $J_8$ 
\begin{equation}
J_{8}\;=\;-\frac{N_{c}}{2\sqrt{3}}B\;=\;-\frac{\sqrt{3}}{2},\;\;\;\;
Y'\;=\;\frac{2}{\sqrt{3}}J_{8}\;=\;-\frac{N_{c}}{3}\;=\;-1,
\label{eq:constraintJ}
\end{equation}
where $B$ represents the baryon number. It is related to the eighth
component of the soliton angular velocity that is due to the presence
of the discrete valence quark level in the Dirac-sea spectrum in the
SU(3) $\chi$QSM~\cite{Blotz:1992pw,Christov:1995vm}, while it 
arises from the Wess-Zumino term in the SU(3) 
Skyrme model~\cite{Witten:1983tx,Guadagnini:1983uv,Jain:1984gp}.
Its presence has no effects on the chiral soliton but allows us to
take only the $\mathrm{SU(3)}$ irreducible representations with zero
triality. Thus, the allowed $\mathrm{SU(3)}$ multiplets are the baryon
octet ($J=1/2$), decuplet ($J=3/2$), and antidecuplet ($J=1/2$),
etc.

The baryon collective wavefunctions of $H$ are written 
as the SU(3) Wigner $D$ functions in representation $\mathcal{R}$:
\begin{eqnarray}
\langle A|\mathcal{R},\, B(Y\, T\, T_{3},\; Y^{\prime}\, J\,
J_{3})\rangle 
& = & \Psi_{(\mathcal{R}^{*}\,;\, Y^{\prime}\, J\,
  J_{3})}^{(\mathcal{R\,};\, 
Y\, T\, T_{3})}(A)\cr
& = & \sqrt{\textrm{dim}(\mathcal{R})}\,(-)^{J_{3}+Y^{\prime}/2}\,
 D_{(Y,\, T,\, T_{3})(-Y^{\prime},\,
 J,\,-J_{3})}^{(\mathcal{R})*}(A),
\label{eq:Wigner}
\end{eqnarray}
 where $\mathcal{R}$ stands for the allowed irreducible representations
of the $\mathrm{SU(3)}$ group, i.e. 
$\mathcal{R}\,=\,8,\,10,\,\overline{10},\cdot\cdot\cdot$
and $Y,\, T,\, T_{3}$ are the corresponding hypercharge, isospin,
and its third component, respectively. The constraint of the right
hypercharge $Y^{\prime}\;=\;1$ selects a tower of allowed $\mathrm{SU(3)}$
representations: The lowest ones, that is, the baryon octet and decuplet,
coincide with those of the quark model. This has been considered as
a success of the collective quantization and as a sign of certain
duality between a rigidly rotating heavy soliton and a constituent
quark model. The third lowest representation is the 
antidecuplet~\cite{Diakonov:1997mm} that includes the $\Theta^+$ and
$N^*$ baryons.  

Different SU(3) representations get mixed in the presence of the
symmetry-breaking term $H_{\mathrm{sb}}$ of the collective Hamiltonian
in Eq.~(\ref{eq:sbH}), so that the collective wave functions are no 
longer in pure states but are given as the following linear
combinations \cite{Blotz:1992pw,Kim:1995mr}:  
\begin{eqnarray}
\left|B_{8}\right\rangle  
& = & \left|8_{1/2},B\right\rangle 
\;+\; c_{\overline{10}}^{B}\left|\overline{10}_{1/2},B\right\rangle 
\;+\; c_{27}^{B}\left|27_{1/2},B\right\rangle ,\cr
\left|B_{10}\right\rangle  & = & \left|10_{3/2},B\right\rangle 
\;+\; a_{27}^{B}\left|27_{3/2},B\right\rangle 
\;+\; a_{35}^{B}\left|35_{3/2},B\right\rangle ,\cr
\left|B_{\overline{10}}\right\rangle  
& = &
\left|\overline{10}_{1/2},B\right\rangle \;+\;
d_{8}^{B}\left|8_{1/2},B\right\rangle \;+\;
d_{27}^{B}\left|27_{1/2},B\right\rangle \;+\;
d_{\overline{35}}^{B}\left|\overline{35}_{1/2},B\right\rangle.
\label{eq:admix}
\end{eqnarray}
The detailed expressions for the coefficients in Eq.(\ref{eq:admix})
can be found in Refs.~\cite{Blotz:1992pw,Ellis:2004uz}.

Since we take into account the effects of isospin symmetry breaking, 
we also have to introduce the EM mass corrections to the mass
splitting of the SU(3) baryons, which are equally important. 
The EM corrections to the baryon masses can
be derived from the baryonic two-point correlation functions. 
The corresponding collective operator was already derived in
Ref.~\cite{Yang:2010id}: 
\begin{equation}
M_{B}^{\mathrm{EM}}\;=\; \langle B|\mathcal{O}^{\mathrm{EM}}|B\rangle,
\label{eq:corr}
\end{equation}
where
\begin{equation}
\mathcal{O}^{\mathrm{EM}} 
\;=\; c^{(1)} D_{\Lambda\Lambda}^{(1)} \;+\; 
\,c^{(8)} \left(\sqrt{3}D_{\Sigma^{0}\Lambda}^{(8)}
+D_{\Lambda\Lambda}^{(8)}\right)
+ c^{(27)} \left(\sqrt{5}D_{\Sigma_{2}^{0}\Lambda_{27}}^{(27)}
+\sqrt{3} D_{\Sigma_{1}^{0}\Lambda_{27}}^{(27)} +
D_{\Lambda_{27}\Lambda_{27}}^{(27)}\right).
\label{eq:emop3}  
\end{equation}
The unknown parameters $c^{(8)}$ and $c^{(27)}$ are determined by the
experimental data for the EM mass splittings of the baryon octet,
while $c^{(1)}$ can be absorbed in the center of baryon masses. The
values of $c^{(8)}$ and $c^{(27)}$ were obtained as 
\begin{equation}
c^{(8)} = -0.15\pm 0.23,\;\;\;\; c^{(27)} = 8.62\pm 2.39   
\end{equation}
in units of MeV~\cite{Yang:2010id}.

The final expressions for the masses of $\Theta^+$ and $N^*$ are
given as 
\begin{eqnarray}
M_{\Theta^+} &=& \overline{M}_{\overline{\bm{10}}} + \frac14
\left(c^{(8)} - \frac4{21} c^{(27)}\right) - 2(m_s - \bar{m}) \delta
,\cr
M_{N^*} &=& \overline{M}_{\overline{\bm{10}}} 
+\frac14\left(c^{(8)} - \frac{32}{63} c^{(27)} \right) T_3 + \frac14
\left(c^{(8)} + \frac{8}{63} c^{(27)} \right) \left(T_3^2 +
  \frac14\right) - (m_{\mathrm{d}}-m_{\mathrm{u}}) T_3 \, \delta -
(m_{\mathrm{s}} - \bar{m})\,\delta,   
\label{eq:masses}
\end{eqnarray}
where $\overline{M}_{\overline{\bm{10}}}$ denotes the center of the
mass splittings of the baryon antidecuplet and $\delta$ is a parameter
defined as 
\begin{equation}
\delta \;=\; -\frac18 \alpha - \beta + \frac1{16}\gamma.  
\label{eq:delta}
\end{equation}

The collective operators for the magnetic moments and axial-vector
constants can respectively be parameterized by six parameters that can
be treated as free~\cite{Kim:1997ip,Kim:1998gt,Kim:1999uf}:  
\begin{eqnarray}
  \label{eq:colmag}
\hat{\mu}  &=&w_{1}D_{X3}^{(8)}\;+\;w_{2}d_{pq3}D_{Xp}^{(8)}\cdot
\hat{J}_{q}\;+\;\frac{w_{3}}{\sqrt{3}}D_{X8}^{(8)}\hat{J}_{3}\cr
&+& \frac{w_{4}}{\sqrt{3}}d_{pq3}D_{Xp}^{(8)}D_{8q}
^{(8)}+w_{5}\left(D_{X3}^{(8)}D_{88}^{(8)}+D_{X8}^{(8)}D_{83}^{(8)}\right)
\;+\;w_{6}\left(D_{X3}^{(8)}D_{88}^{(8)}-D_{X8}^{(8)}D_{83}^{(8)}\right),\cr
\hat{g}_A &=&a_1 D_{X3}^{(8)} \;+\; a_2d_{pq3}D_{Xp}^{(8)}\,\hat{J}_q
\;+\; \frac{a_3}{\sqrt{3}} D_{X8}^{(8)}\,\hat{J}_3 \cr 
&+& \frac{a_4}{\sqrt{3}}d_{pq3} D_{Xp}^{(8)}\,D_{8q}^{(8)} \;+ \;a_5\left(
D_{X3}^{(8)}\,D_{88}^{(8)}+D_{X8}^{(8)}\,D_{83}^{(8)}\right)
\cr 
&+& a_6\left(D_{X3}^{(8)}\,D_{88}^{(8)}-D_{X8}^{(8)}\,D_{83}^{(8)}\right), 
\label{Eq:g1}
\end{eqnarray}
where $\hat{J}_q$ ($\hat{J}_3$) stand for the $q$-th (third)
component of the spin operator of the baryons. 
The parameters $w_i$ and $a_i$ can be unambiguously fixed by using the
magnetic moments and semileptonic decay constants of the baryon 
octet~\cite{Yang,YangKim}. We refer to Refs.~\cite{Yang} for the
detailed expressions for the $N^*\to N$ transition magnetic moments
and for the $\Theta^+$ magnetic moment and axial-vector constants for
the $\Theta\to KN$ decay. 

\section{Results and discussion}
In order to find the masses of the baryon antidecuplet, we
need to fix the relevant parameters. There are several ways to fix
them. For example, Diakonove et al.~\cite{Diakonov:1997mm} use the 
mass splittings of the baryon octet and decuplet, $\pi N$ sigma term, and
the $N^*$ mass that was then taken to be around 1710 MeV. The $\pi N$
sigma term was taken from Ref.~\cite{Gasser:1990ce}, i.e. $\sigma_{\pi
  N} \approx 45$ MeV. In addition, the ratio of the curren quark mass
$m_{\mathrm{s}} /(m_{\mathrm{u}}+m_{\mathrm{d}})\approx 12.5$ was
quoted from Ref.~\cite{Leutwyler:1996qg} to determine the parameters:
\begin{equation}
m_{\mathrm{s}} \alpha\approx -218\,\mathrm{MeV}, \;\;\;\; m_{\mathrm{s}}
\beta\approx-156\,\mathrm{MeV}, \;\;\;\;
m_{\mathrm{s}} \gamma\approx-107\,\mathrm{MeV}. 
  \label{eq:DPP}
\end{equation}
On the other hand, Ellis et al.~\cite{Ellis:2004uz} carried out the
analysis for the mass splittings of the baryon 
antidecuplet, based on the then experimental data of the $\Theta^+$
and $\Xi^{--}$ masses together with those of the baryon octet and
decuplet. They predicted the $\pi N$ sigma term $\sigma_{\pi N} =
73\, \mathrm{MeV}$ from the fitted values of the parameters:
\begin{equation}
I_2=0.49\,\mathrm{fm},\;\;\;\;
m_{\mathrm{s}} \alpha=-605\,\mathrm{MeV}, \;\;\;\;
m_{\mathrm{s}} \beta=-23\,\mathrm{MeV}, \;\;\;\;
m_{\mathrm{s}}\gamma=152\,\mathrm{MeV}.
  \label{eq:Ellis}  
\end{equation}
Very recently, Ref.~\cite{Yang:2010fm}
reanalyzed the mass splittings of the SU(3) baryons within a $\chi$SM,
employing isospin symmetry breaking. An obvious advantage of
including the effects of isospin symmetry breaking is that one can
fully utilize the whole experimental data of the octet masses to fix
the parameters. Using the baryon octet masses, $\Omega^-$ mass
$(1672.45\pm0.29)\,\mathrm{MeV}$~\cite{PDG}, and $\Theta^+$ mass
$(1524\pm 5)\,\mathrm{MeV}$~\cite{Nakano:2008ee}, both of which
are the isosinglet baryons   
the key parameters were found to be: 
\begin{equation}
I_2=(0.420\pm 0.006)\,\mathrm{fm}, \;\;\;\;
m_{\mathrm{s}} \alpha=(-262.9\pm5.9)\,\mathrm{MeV}, \;\;\;\;
m_{\mathrm{s}} \beta=(-144.3\pm3.2)\,\mathrm{MeV},\;\;\;\;
m_{\mathrm{s}} \gamma=(-104.2\pm2.4)\,\mathrm{MeV}. 
  \label{eq:present}  
\end{equation}
In addition, the $\pi N$ sigma term was predicted as $\sigma_{\pi
  N}=(36.4\pm3.9)\,\mathrm{MeV}$. Since $\delta$ defined in
Eq.(\ref{eq:delta}), let us compare its values from each work
mentioned above. The corresponding results are given, respectively, as
follows: 
\begin{equation}
m_{\mathrm{s}} \delta \;=\; 177\,\mathrm{MeV}~(\mbox{Diakonov et al.}),
\;\;\; 
m_{\mathrm{s}} \delta\;=\; 108\,\mathrm{MeV}~(\mbox{Ellis et al.}),
\;\;\; 
m_{\mathrm{s}} \delta \;=\; 171\,\mathrm{MeV}~(\mbox{present work}),
\end{equation}
with isospin symmetry breaking switched off. 
If we use the LEPS experimental data~\cite{Nakano:2008ee} for
$M_{\Theta^+}$, we can immediately obtain the corresponding masses of
the $N^*$, respectively:
\begin{equation}
M_{N^*} \;=\; 1700\,\mathrm{MeV}~(\mbox{Diakonov et al.}),
\;\;\; 
M_{N^*} \;=\; 1631\,\mathrm{MeV}~(\mbox{Ellis et al.}),
\;\;\; 
M_{N^*} \;=\; 1694\,\mathrm{MeV}~(\mbox{present work}),
  \label{eq:Nstar}  
\end{equation}
If one employs the DIANA data~\cite{Barmin:2009cz}, the $N^*$ mass is
yielded as  
\begin{equation}
M_{N^*} \;=\; 1715\,\mathrm{MeV}~(\mbox{Diakonov et al.}),
\;\;\; 
M_{N^*} \;=\; 1646\,\mathrm{MeV}~(\mbox{Ellis et al.}),
\;\;\; 
M_{N^*} \;=\; 1708\,\mathrm{MeV}~(\mbox{present work}). 
\end{equation}
The comparison made above already indicates that the predicted masses
of the $N^*(1685)$ resonance from the previous analyses are deviated
from the experimental data. Moreover, it is essential to take into
account the effects of isospin symmetry breaking, in order to produce
the mass of the $N^*$ resonance quantitatively~\cite{Yang:2010fm}.
Since there are, however, two different experimental values of the
$\Theta^+$ mass from the LEPS and DIANA collaborations,
it is necessary to examine carefully the dependence of the relevant
observables on that of the $\Theta^+$ baryon rather than choosing one
specific value of $M_{\Theta^+}$ to fit the parameters. Thus, in the
present Section, we discuss the dependence of relevant observables on
$M_{\Theta^+}$, taking it as a free parameter.

\begin{figure}[ht]
  \centering
\includegraphics[scale=0.6]{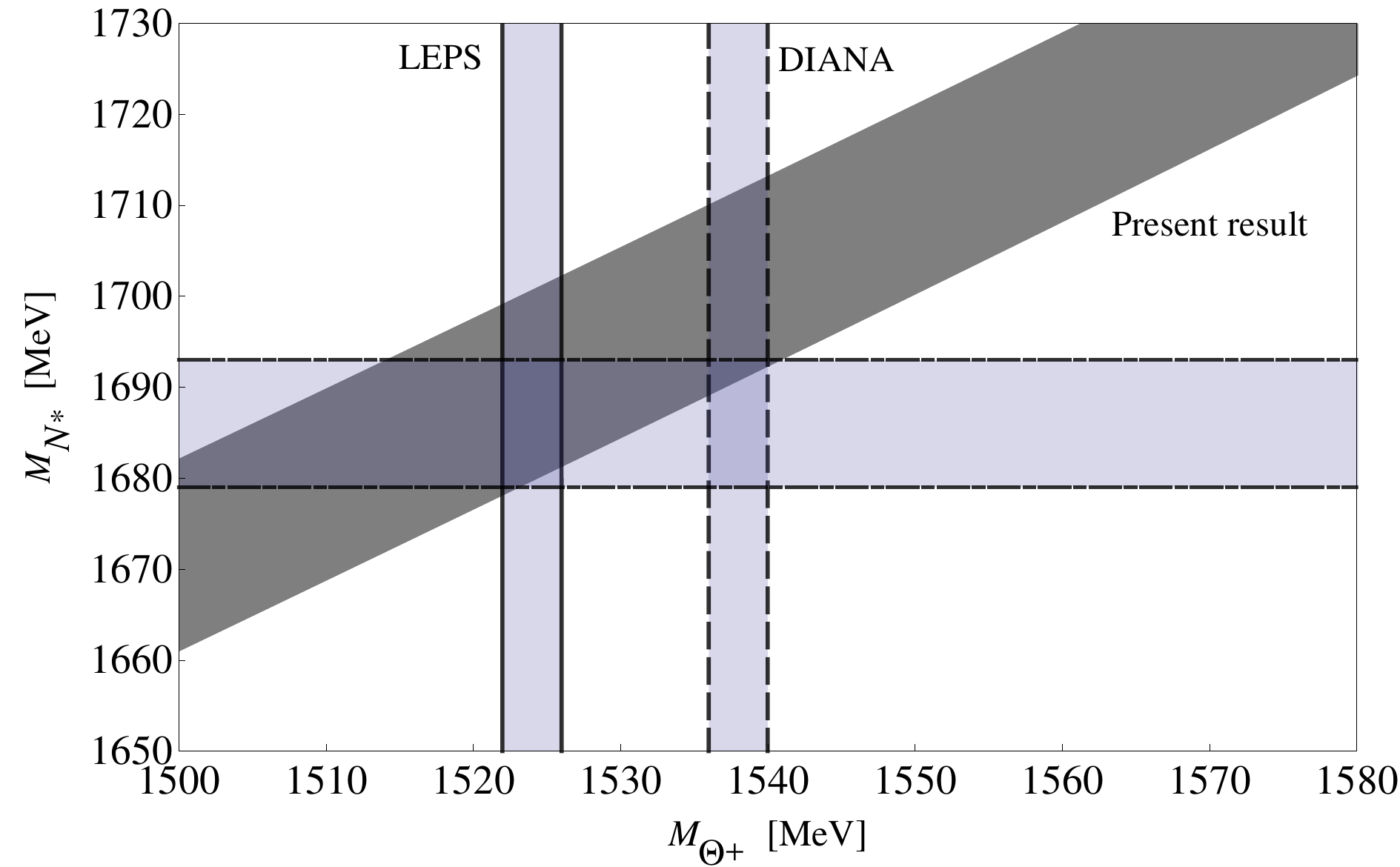}  
  \caption{The dependence of the $N^*$ mass on $M_{\Theta^+}$. The
    vertical shaded bars bounded with the solid and 
dashed lines denote the measured values of the $\Theta^+$ mass with
uncertainties by the LEPS and DIANA collaborations, respectively. The
horizontal shaded region draws the values of the $N^*$ mass with
uncertainty taken from Ref.~\cite{Kuznetsov:2008ii}. The sloping shaded
region represents the present results of the $M_{\Theta^+}$ dependence
of the $N^*$ mass.}
  \label{fig:1}
\end{figure}
In Fig.~\ref{fig:1}, we draw the $N^*$ mass as a function of
$M_{\Theta^+}$. The vertical shaded bars bounded with the solid and
dashed lines denote the measured values of the $\Theta^+$ mass with
uncertainties by the LEPS and DIANA collaborations, respectively. The
horizontal shaded region denotes the values of the $N^*$ mass with
uncertainty taken from Ref.~\cite{Kuznetsov:2008ii}. The sloping shaded
region shows the dependence of the $N^*$ mass on 
$M_{\Theta^+}$. The $N^*$ mass increases monotonically, as
$M_{\Theta^+}$ increases. This behavior can be easily understood from
Eq.(\ref{eq:masses}): the mass of the $N^*$ resonance depends linearly
on the parameter $\delta$.
Interestingly, if we take the $M_{\Theta^+}$ value of the LEPS
experiment, i.e., $M_{\Theta^+}=1524\,\mathrm{MeV}$, we obtain 
$M_{N^*}\simeq1690\,\mathrm{MeV}$, which is in good agreement with the 
experimental data:
$M_{N^*}=(1685\pm12)\,\mathrm{MeV}$~\cite{Kuznetsov:2008ii}.   
On the other hand, if we use the value of $M_{\Theta^+}$ measured by
the DIANA collaboration, the $N^*$ mass turns out to be larger than
$1690$ MeV.  It implies that the $\Theta^+$ mass reported by the LEPS
collaboration~\cite{Nakano:2008ee} is consistent with that of
$N^*(1685)$ from recent experiments~\cite{Kuznetsov:2006kt,
CBELSA,Jaegle:2008ux,Jaegle:2011sw,Kuznetsov:2011zz}, at least, at the
present framework of a $\chi$SM with isospin symmetry
breaking~\cite{Yang:2010fm}.  

The parameters $w_i$ in Eq.(\ref{eq:colmag}) can be fitted by the
magnetic moments of the baryon
octet~\cite{Kim:1997ip,Kim:1998gt,Kim:1999uf}. However, since the
mixing coefficients appearing in Eq.~(\ref{eq:admix}) depend
explicitly on $\alpha$ and $\gamma$, the parameters $w_i$ are also
given as functions of $\sigma_{\pi N}$ through $\alpha$ and
$\gamma$ as shown in Ref.~\cite{Kim:2005gz}. As previously
mentioned, since the mass parameters $\alpha$ and $\gamma$ as well as
$\sigma_{\pi N}$ were unambiguously fixed in Ref.~\cite{Yang:2010fm},
we can derive the transition magnetic moments for the $N^*\to N\gamma$
decay unequivocally. Explicitly, the transition magnetic moments 
$\mu_{pp^*}$ and $\mu_{nn^*}$ are recapitulated, 
respectively, as follows ~\cite{Kim:2005gz}:
\begin{eqnarray}
\mu_{pp^{\ast}}^{(0)} 
& = & 0,\cr
\mu_{pp^{\ast}}^{(\mathrm{op})} 
& = & -\frac{1}{27\sqrt{5}}w_{4}-\frac{1}{18\sqrt{5}}\left(w_{5}
+\frac{3}{2}w_{6}\right),\cr
\mu_{pp^{\ast}}^{(\mathrm{wf})} 
& = & -\frac{5}{24\sqrt{5}}\left(w_{1}
+\frac{5}{2}w_{2}-\frac{1}{2}w_{3}\right)c_{\overline{10}}
-\frac{35}{72\sqrt{5}}\left(w_{1}-\frac{11}{14}w_{2}
-\frac{3}{14}w_{3}\right)c_{27}\cr
 &  &
 +\left[\frac{1}{2\sqrt{5}}\left(w_{1}
-\frac{1}{2}w_{2}+\frac{1}{6}w_{3}\right)
-\frac{7}{6\sqrt{5}}\left(w_{1}-\frac{1}{2}w_{2}
-\frac{1}{14}w_{3}\right)\right]d_{8}\cr
 &  &
 -\frac{1}{45\sqrt{5}}\left(w_{1}+2w_{2}
-\frac{3}{2}w_{3}\right)d_{27}, \cr
\mu_{nn^{\ast}}^{(0)} 
& = & \frac{1}{6\sqrt{5}}\left(w_{1}+w_{2}+\frac{w_{3}}{2}\right),\cr
\mu_{nn^{\ast}}^{(\mathrm{op})} 
& = & -\frac{1}{54\sqrt{5}}w_{4}
+\frac{1}{18\sqrt{5}}\left(w_{5}+\frac{3}{2}w_{6}\right),\cr
\mu_{nn^{\ast}}^{(\mathrm{wf})} 
& = &
\frac{7}{36\sqrt{5}}\left(w_{1}-\frac{11}{14}w_{2}-\frac{3}{14}w_{3}\right)c_{27}
+\frac{1}{2\sqrt{5}}\left(w_{1}-\frac{1}{2}w_{2}+\frac{1}{6}w_{3}\right)d_{8}\cr
 &  &
 -\frac{1}{90\sqrt{5}}\left(w_{1}+2w_{2}-\frac{3}{2}w_{3}\right)d_{27}.
\label{eq:mu_nnstar}
\end{eqnarray}
As already discussed in Ref.~\cite{Kim:2005gz}, $\mu_{pp^*}$ vanishes
in the SU(3) symmetric case. Thus, $\mu_{pp^*}$ is only finite with
the effects of SU(3) symmetry breaking included. 

\begin{figure}[ht]
  \centering
\includegraphics[scale=0.6]{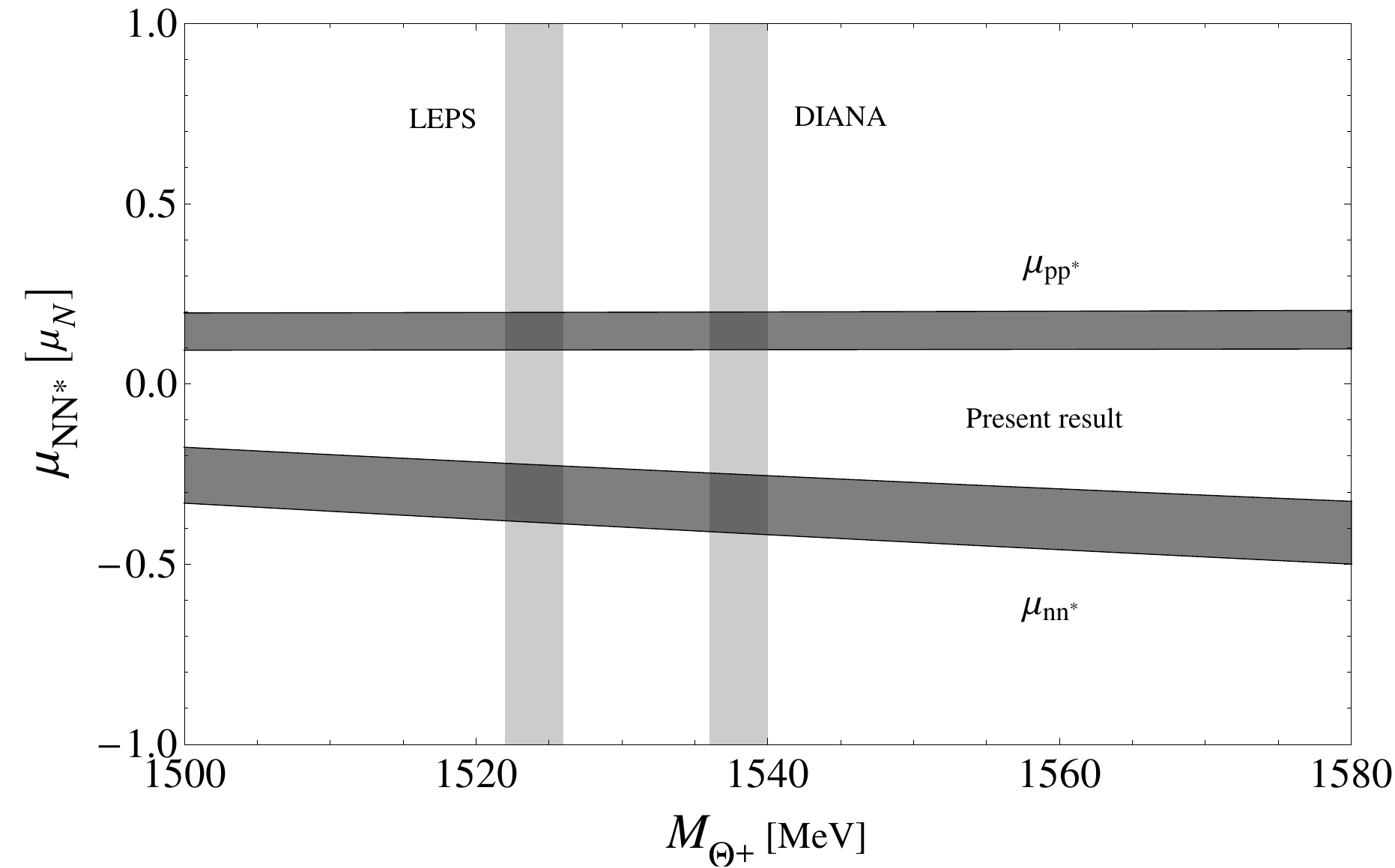}  
  \caption{The dependence of the transition magnetic moments for the
    $N^*\to N\gamma$ decay on $M_{\Theta^+}$. The
    vertical shaded bars bounded with the solid and 
dashed lines denote the measured values of the $\Theta^+$ mass with
uncertainties by the LEPS and DIANA collaborations, respectively. The
horizontal shaded regions stands for the present results of the
$M_{\Theta^+}$ dependence of the transition magnetic moments
$\mu_{p^*p}$ and $\mu_{n^*n}$.}
  \label{fig:2}
\end{figure}
Figure~\ref{fig:2} shows the transition magnetic moments for the
$N^*\to N\gamma$ decay as functions of $M_{\Theta^+}$. While
$\mu_{pp^*}$ is almost independent of $M_{\Theta^+}$, $\mu_{nn^*}$  
decreases slowly as $M_{\Theta^+}$ increases.  
On the other hand, the magnitude of $\mu_{nn^*}$ turns out 
to be larger than that of $\mu_{pp^*}$ as was already pointed out in 
Refs.~\cite{Polyakov:2003dx,Azimov:2005jj,Kim:2005gz}. 

\begin{table}[h]
  \centering
\caption{The results of the $N^*\to N$ transition magnetic moments in
  units of the nuclear magneton $\mu_N$ and of the radiative decay
  widths in unit of keV. The mass $M_{\Theta^+}=(1524\pm5)$ MeV is
used as an input.}
  \begin{tabular}{c|crrrc}
    \hline \hline
$\mu_{NN^*}$ 
& $\mu_{NN^*}^{(0)}$
&
$\mu_{NN^*}^{(\mathrm{op})}\;\;\;\;$
&
$\mu_{NN^*}^{(\mathrm{wf})}\;\;\;\;$
& $\mu_{NN^*}^{(\mathrm{total})}\;\;\;\;$
& $\Gamma_{NN^*}\,[\mathrm{keV}]$
\\
\hline 
$\mu_{pp^{\ast}}$ 
& $0$ 
& $0.272\pm0.051$ 
& $-0.125\pm0.013$ 
& $0.146\pm0.053$ 
& $17.7\pm3.2$\\
$\mu_{nn^{\ast}}$ 
& $-0.252\pm0.077$ 
& $-0.159\pm0.042$ 
& $0.107\pm0.003$ 
& $-0.304\pm0.089$ 
& $77.1\pm11.3$\\
\hline \hline
  \end{tabular}
    \label{tab:1}
\end{table}
In Table~\ref{tab:1}, we list each contribution to $\mu_{NN^*}$ as
well as the radiative decay widths for $N^*\to N\gamma$ with the mass
of the $\Theta^+$ from the LEPS experiment used. 
Note that the sign of $\mu_{nn^*}$ is negative whereas that of
$\mu_{pp^*}$ is positive. However, the previous result for
$\mu_{nn^*}$ was positive ~\cite{Kim:2005gz}. The reason can be found
in the different values of $w_i$. Let us compare closely  the present
results with those of Ref.~\cite{Kim:2005gz}, considering only the
SU(3) symmetric part without loss of generality. In fact, $w_i$ derived in 
Ref.~\cite{Kim:2005gz} depends on $\sigma_{\pi N}$:
\begin{equation}
w_1^{\mathrm{old}} \;=\; -3.736-0.107\, \sigma_{\pi N},\;\;\;
w_2^{\mathrm{old}} \;=\; 24.37-0.21 
\,\sigma_{\pi N}, \;\;\; w_3^{\mathrm{old}} \;=\; 7.547. 
\end{equation}
If one takes the value of the $\pi N$ sigma terms as $\sigma_{\pi N}
\approx 40$ MeV (70 MeV), one gets
\begin{equation}
w_1^{\mathrm{old}} \;=\; -8.14\,(-11.44),\;\;\;w_2^{\mathrm{old}} \;=\;
15.97\,(9.67),\;\;\; w_3^{\mathrm{old}} \;=\; 7.547,   
\end{equation}
while the results in this work use the newly obtained values of
$w_i$~\cite{YangKim}  
\begin{equation}
 w_1 \;=\; -12.95\pm0.10,\;\;\; w_2 \;=\; 5.388\pm 0.933, 
\;\;\; w_3 \;=\; 8.354\pm0.861.
\end{equation}
Thus, the magnitude of the present $w_1$ is larger than those of
$w_1^{\mathrm{old}}$, whereas that of $w_2$ turns out to be smaller
than those of $w_ 2^{\mathrm{old}}$. Since $w_1$ and $w_2$ have
different sign as in Eq.(\ref{eq:mu_nnstar}), they destructively
interfere eath other, so that the sign of $\mu_{nn^*}$ 
becomes negative in the present case but it is positive in
Ref. ~\cite{Kim:2005gz}. However, magnetic properties of the octet
and decuplet baryons are almost intact because of the constructive
interference of $w_1$ and $w_2$, even though we have the
different values of $w_i$. 
The ratios of the transition magnetic moments and of the radiative
decay widths are obtained as 
\begin{equation}
\left|\frac{\mu_{nn^{\ast}}}{\mu_{pp^{\ast}}}\right| 
\;=\; 2.08\pm0.97,\;\;\;\;
\frac{\Gamma_{nn^{\ast}}}{\Gamma_{pp^{\ast}}} 
\;=\; 4.36\pm1.02.
\end{equation}

\begin{figure}[ht]
  \centering
\includegraphics[scale=0.6]{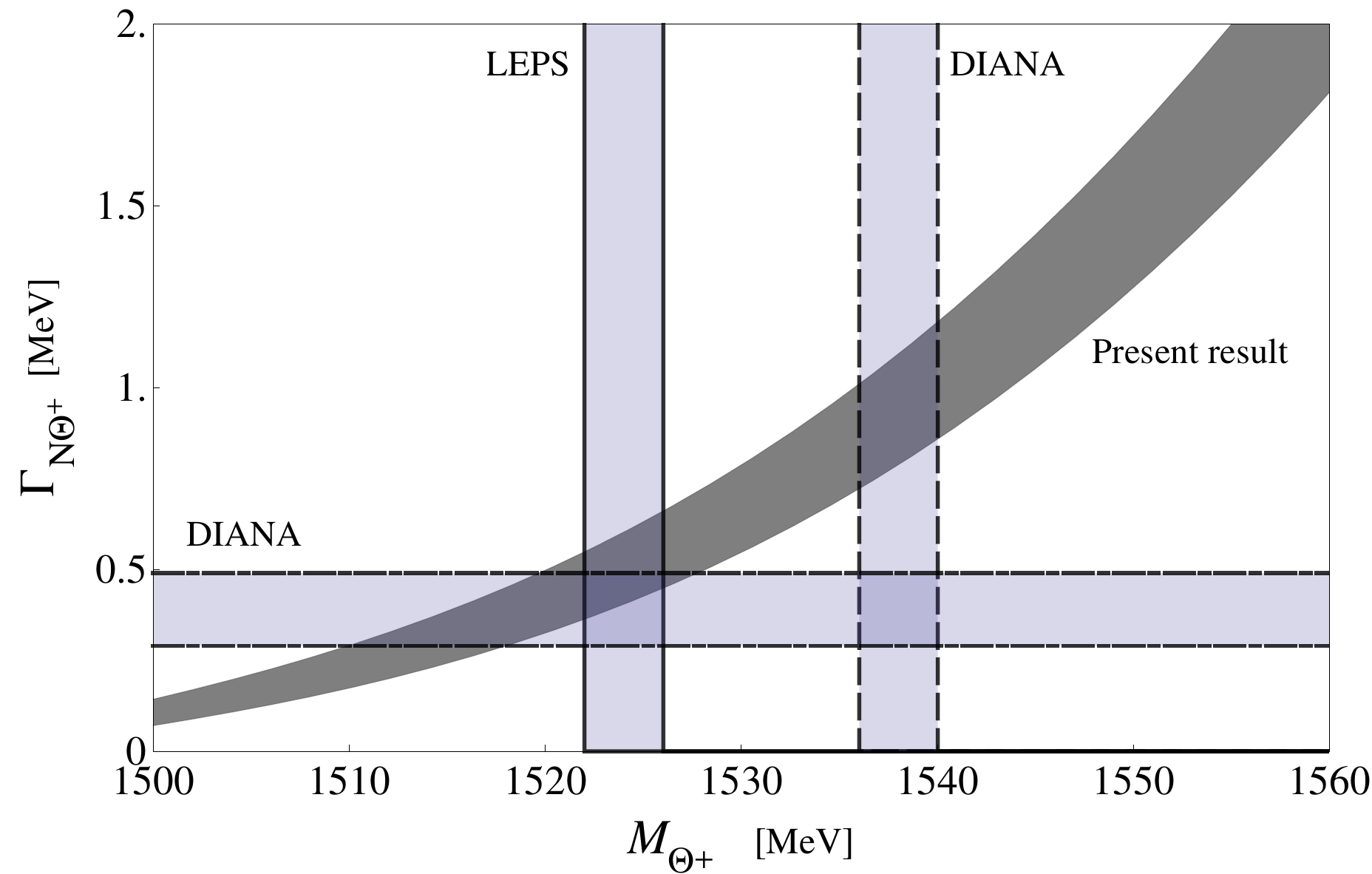}  
  \caption{The dependence of the decay width $\Gamma_{N\Theta^+}$ for
the $\Theta^+\to KN$ decay on $M_{\Theta^+}$. The vertical shaded bars
bounded with the solid and dashed lines denote the measured values of
the $\Theta^+$ mass with uncertainties by the LEPS and DIANA
collaborations, respectively. The horizontal shaded region draws the
values of the $N^*$ mass with uncertainty taken from
Ref.~\cite{Kuznetsov:2008ii}. The sloping shaded 
region represents the present results of the $M_{\Theta^+}$ dependence
of $\Gamma_{N\Theta^+}$.}
  \label{fig:3}
\end{figure}
The narrowness of the decay width is one of the peculiar
characteristics of the pentaquark baryons. For example, the decay
width of the $\Theta^+\to KN$ vanishes in the nonrelativistic
limit~\cite{Diakonov:2004ai}. The decay width $\Gamma_{\Theta N K}$
was already studied in chiral soliton models with SU(3) symmetry
breaking taken into account. We refer to
Refs.~\cite{Yang:2007yj,Ledwig:2008rw} for details.
In Fig.~\ref{fig:3}, we examine the dependence of the decay width
$\Gamma_{N\Theta}$ for $\Theta^+\to KN$ on the $\Theta^+$ mass. Being
different from the $N^*$ mass and the transition magnetic moments, the
decay width $\Gamma_{\Theta^+N}$ increases almost quadratically as
$M_{\Theta^+}$ increases. This can be understood from the fact that
the decay width is proportional to the square of the $g_{NK\Theta^+}$
coupling constant which depends linearly on $M_{\Theta^+}$. When
the $\Theta^+$ mass is the same as the value measured by the LEPS
collaboration, $\Gamma_{N\Theta^+}$ turns out to be about 0.5
MeV. However, at the value $M_{\Theta^+}\approx 1540$ MeV
corresponding to that of the DIANA experiment, the decay width
$\Gamma_{N\Theta^+}$ is close to 1 MeV. We want to emphasize that the
decay width of the $\Theta^+$ is still below 1 MeV in the range of
$M_{\Theta^+}$: $1520-1540$ MeV.  When we use the measured value of
$M_{\Theta^+}$ by the LEPS collaboration, we obtain
$\Gamma_{\Theta^+\to NK} = 0.5\pm 0.1$ MeV.  

\begin{figure}[ht]
  \centering
\includegraphics[scale=0.6]{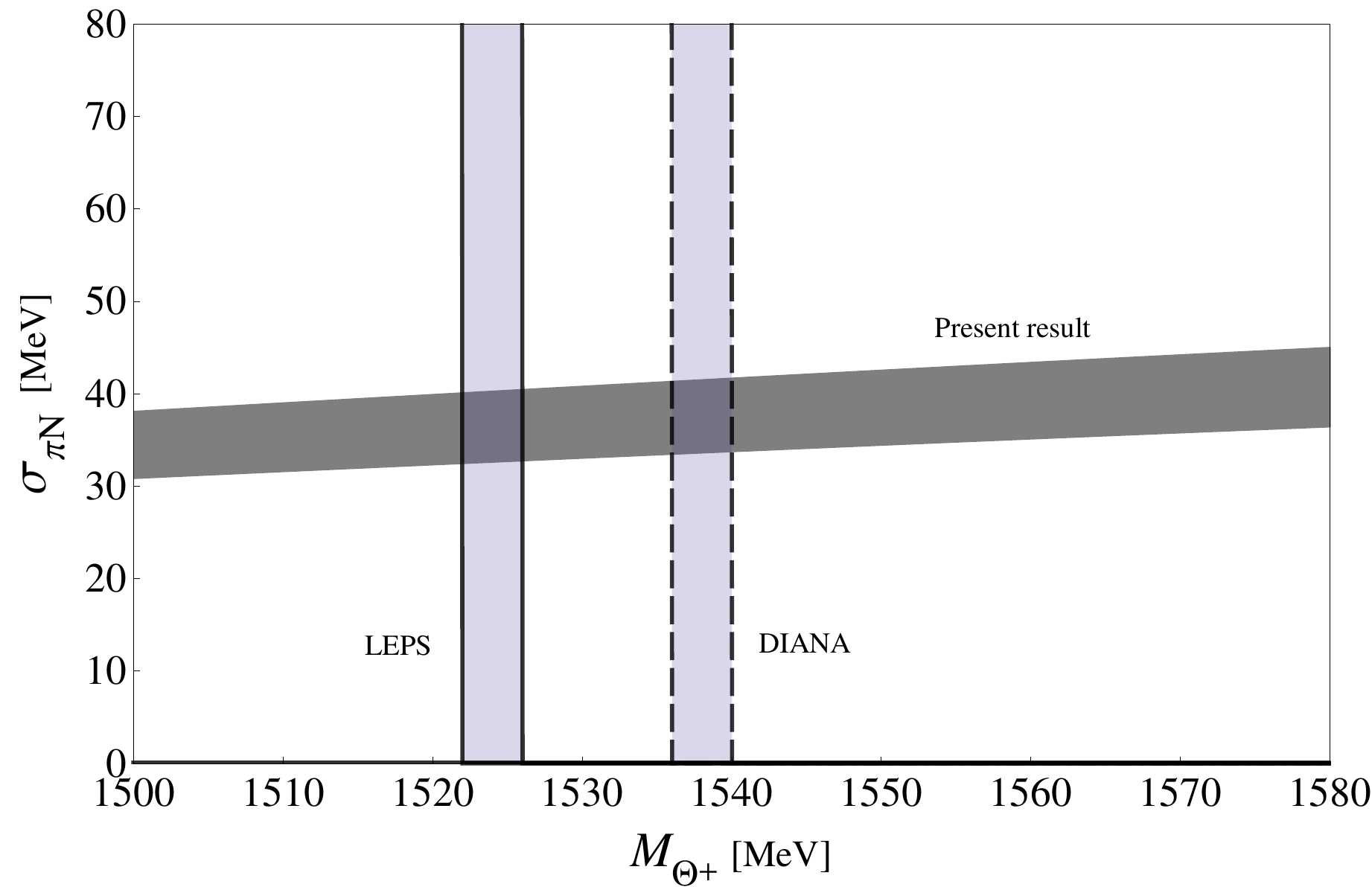}  
  \caption{The dependence of the $\sigma_{\pi N}$ on the $\Theta^+$
mass. The vertical shaded bars bounded with the solid and dashed lines
denote the measured values of the $\Theta^+$ mass with uncertainties
by the LEPS and DIANA collaborations, respectively. The sloping shaded 
region represents the present results of the $M_{\Theta^+}$ dependence
of $\sigma_{\pi N}$.}
  \label{fig:4}
\end{figure}
Figure~\ref{fig:4} depicts predicted values of the $\pi N$ sigma term
as a function of the $\Theta^+$ mass. At first sight, the result is
rather surprising. Firstly, it is almost insensitive to the $\Theta^+$
mass. Secondly, the value of $\sigma_{\pi N}$ is pretty smaller than
those known from the previous works on the baryon
antidecuplet~\cite{Ellis:2004uz,Schweitzer:2003fg}.  In order to
understand the reason of this difference, we want to examine in detail
the $\pi N$ sigma term in comparison with those discussed in previous
works, in particular, with Ref.~\cite{Schweitzer:2003fg}, where the
$\pi N$ sigma term was extensively studied within the same framework.  
Since $\sigma_{\pi N}$ is expressed as
\begin{equation}
\sigma_{\pi N} \;=\; -3 \bar{m} (\alpha + \beta),
  \label{eq:sigma}  
\end{equation}
we need to scrutinize the dependence of $\bar{m}\alpha$ and
$\bar{m}\beta$ on $M_{\Theta^+}$.
\begin{figure}[ht]
  \centering
\includegraphics[scale=0.6]{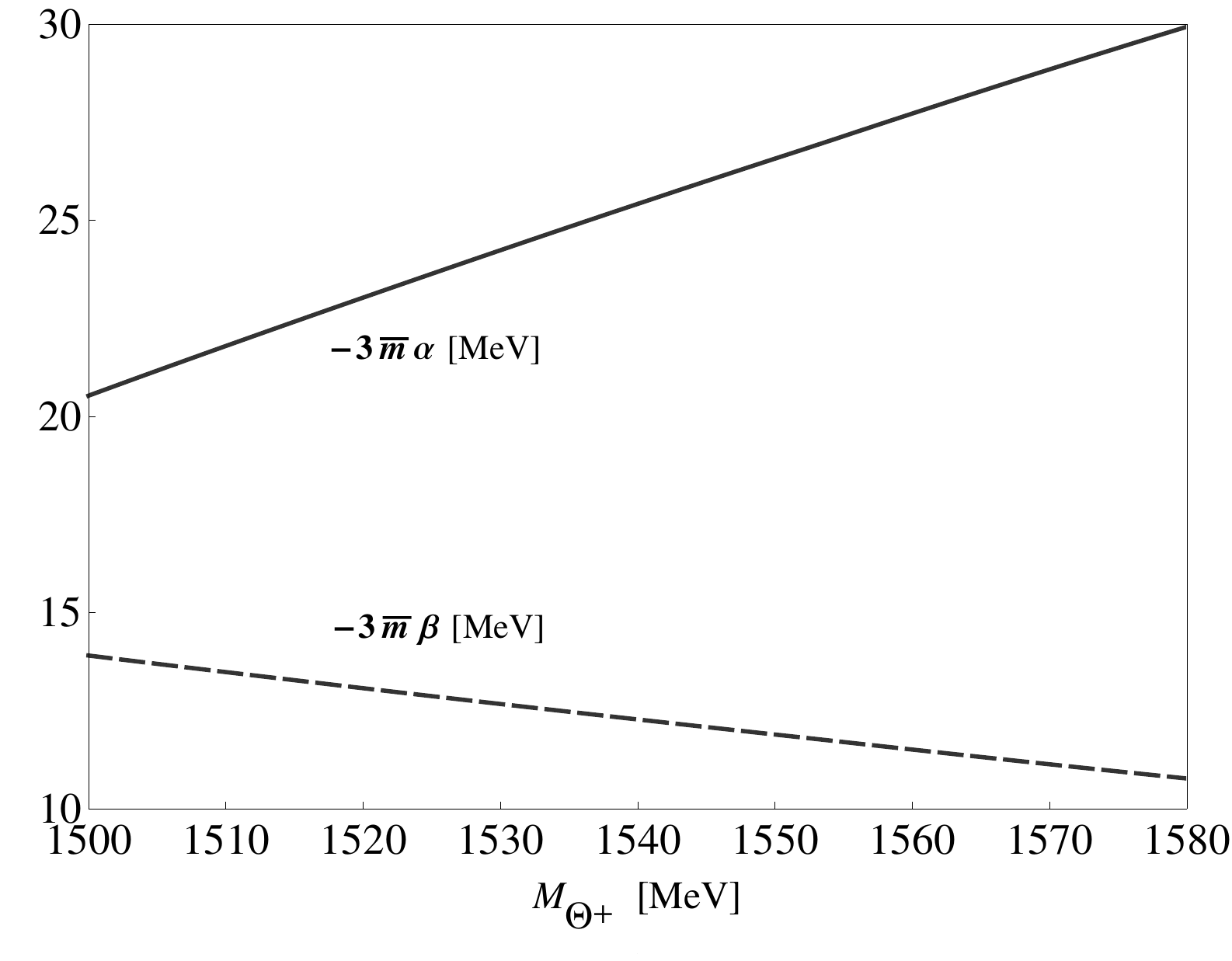}  
  \caption{The results of the parameters $-3\bar{m}\alpha$ and
    $-3\bar{m}\beta$ as functions of $M_{\Theta^+}$.}
  \label{fig:5}
\end{figure}
Figure~\ref{fig:5} depicts the results of the parameters
$-3\bar{m}\alpha$ and $-3\bar{m}\beta$ as functions of $M_{\Theta^+}$.  
Interestingly, while $-3\bar{m}\alpha$ increases monotonically as
$M_{\Theta^+}$ increases, $-3\bar{m}\beta$ decareases almost at the
same rate as $-3\bar{m}\alpha$. Consequently, $\sigma_{\pi N}$ remains 
rather stable. On the other hand, Schweitzer~\cite{Schweitzer:2003fg}
expressed $\sigma_{\pi N}$ in terms of the mass splittings of each
representation: 
\begin{equation}
\frac{m_{\mathrm{s}}}{\bar{m}}  \sigma_{\pi N} \;=\; 3(4 M_{\Sigma} -
3 M_\Lambda - M_N) + 4(M_\Omega - M_\Delta) -4(M_{\Xi_{3/2}} -
  M_{\Theta^+} ).
  \label{eq:Schweit}
\end{equation}
and determined it to be $\sigma_{\pi N}=(74\pm 12)\,\mathrm{MeV}$,
taking the experimental values of $M_{\Theta^+}=1540\,\mathrm{MeV}$
and $M_{\Xi_{3/2}}=1862$ MeV~\cite{Alt:2003vb} for granted at that
time, and using the ratio of the current quark mass
$m_{\mathrm{s}}/\bar{m} =25.9$.  However, using the predicted value of 
$M_{\Xi_{3/2}}\approx 2020\,\mathrm{MeV}$ in Ref.~\cite{Yang:2010fm},
we get $\sigma_{\pi  N}\approx 45\,\mathrm{MeV}$.  Thus, the present
result is not in contracdiction with that of
Ref.~\cite{Schweitzer:2003fg}. 

Taking the effects of isospin symmetry breaking into account,
however, we can rewrite $\sigma_{\pi N}$ in terms of the mass
splittings of the isospin multiplets 
\begin{equation}
\sigma_{\pi N} \;=\; \frac{3\bar{m}}{m_{\mathrm{d}} -
  m_{\mathrm{u}}}\left[\frac{10}{3}(M_{\Sigma^0} - M_{\Sigma^+}) +
  \frac53(M_{\Xi^-} - M_{\Xi^0}) - 4(M_{n^*} - M_{p^*}) \right].     
\label{eq:sigso}
\end{equation}
Plugging the ratio $(m_{\mathrm{d}} - m_{\mathrm{u}})/(m_{\mathrm{u}} +
m_{\mathrm{d}}) =0.28\pm 0.03$~\cite{Gasser:1982ap} into
Eq.(\ref{eq:sigso}), considering the experimental data for the 
corresponding baryon octet masses~\cite{PDG}, and using the values of
the $M_{n^*}$ and $M_{p^*}$ predicted in Ref.~\cite{Yang:2010fm}, we
obtain $\sigma_{\pi N} \approx 34\,\mathrm{MeV}$, which is almost the
same as that of Ref.~\cite{Yang:2010fm}. 

\section{Summary and conclusion}
In the present work, we aimed at investigating various observables of
the baryon antidecuplet $\Theta^+$ and $N^*$, emphasizing on their
dependence on the $\Theta^+$ mass within a chiral soliton 
model. We utilized the mass parameters $\alpha$, $\beta$, and $\gamma$ 
derived unequivocally in Ref.~\cite{Yang:2010fm}. 
We first compared the present
result of the $N^*$ mass with those predicted by the previous 
analyses~\cite{Diakonov:1997mm,Ellis:2004uz}.
We then examined the dependence of the $N^*$ mass on the $\Theta^+$
one. We found that the measured value of the $\Theta^+$ mass by the
LEPS collaboration turned out to be consistent with that of the $N^*$
mass by Kuznetsov and Polyakov~\cite{Kuznetsov:2008ii} within the
present framework. We then scrutinized the transition magnetic moments
of the radiative decay $N^*\to N\gamma$. While $\mu_{pp^*}$ is almost
independent of the $\Theta^+$ mass, $\mu_{nn^*}$ decreases
slowly as $M_{\Theta^+}$ increases. We also discussed the 
results of the $N^*\to N\gamma$ transition magnetic moments with those
of previous works. The decay width of the $\Theta^+$ was studied
and was found to be $0.5\pm 0.1$ MeV when the LEPS data of
$M_{\Theta^+}$ was employed, which is compatible with the
corresponding measured decay width by the DIANA collaboration.   
Finally, we analyzed the $\pi N$ sigma term within the present
framework. It turned out that $\sigma_{\pi N}$ was almost independent
of the $\Theta^+$ mass. We explained the reason why it was rather 
smaller than those in previous analyses, in particular, in
Ref.~\cite{Schweitzer:2003fg}. In addition, we found a new expression
for the $\pi N$ sigma term in terms of the isospin mass splittings of
the hyperon octet as well as that of the antidecuplet $N^*$.  

\begin{acknowledgments}
We are grateful to T. Nakano for suggesting the analysis of the
$\Theta^+$ mass dependence of relevant observables. The present
work was supported by Basic Science Research Program through the
National Research Foundation of Korea (NRF) funded by the Ministry of
Education, Science and Technology (grant number: 2010-0016265).   
\end{acknowledgments}
%\cite{Jezabek:1987ns}

\end{document}